\documentclass[conference]{IEEEtran}

\usepackage{subfigure}

\usepackage[noadjust]{cite}

\ifCLASSINFOpdf
   \usepackage[pdftex]{graphicx}
\else
  \usepackage[dvips]{graphicx}
\fi

\usepackage[cmex10]{amsmath}
\usepackage{fixltx2e}

\usepackage{float}
\usepackage{amsmath}
\begin{document}
%
\title{Machine Learning Based Featureless Signalling}

\author{\IEEEauthorblockN{Ismail Shakeel}
\IEEEauthorblockA{Defence Science and Technology Group \\ 
Edinburgh, SA, Australia\\
Email: ismail.shakeel@dst.defence.gov.au}
}

\maketitle

\begin{abstract}
Direct-sequence spread-spectrum (DSSS) is commonly used to mitigate the effect of jamming and to operate under an adversary receiver's thermal noise floor in order to avoid signal detection.  Unfortunately, the discrete nature and unique distribution of DSSS spreading sequences make it relatively easy to detect the resulting transmitted signals.  To overcome this issue, this paper proposes a machine learning based scheme that generates featureless, non-repetitive noise-like spread signals. 
The proposed scheme provides several benefits over the standard DSSS system including the ability to generate signals with low probabilities of detection/intercept, additional processing gain and also an uncoordinated synchronisation method.

\end{abstract}


\IEEEpeerreviewmaketitle


\section{Introduction}

Machine Learning (ML) has been successfully applied in many fields including wireless communications.  Recently,  several ML-based schemes have been proposed for the physical layer \cite{tom_dec17,wang_nov17}.  One of these proposed schemes \cite{tom_dec17} views communication systems as an end-to-end reconstruction optimisation task and jointly optimises both transmitter and receiver components against a performance target for a given channel environment.  This paper takes a similar ML approach and proposes a Jamming-Resilient (JR) signalling scheme to generate communication signals that have Low Probabilities of Detection and Intercept (LPD/LPI).  LPD/LPI features of a communication waveform make it difficult for an unintended receiver to detect and/or extract useful features from transmissions.  These extracted features could be used by an adversary to determine the geographical location of the transmitter, exploit system's communication protocols or jam transmissions.  Features from LPD/LPI signals are extracted using methods that use advanced signal processing techniques such as time-frequency transforms, higher-order moments, cyclostationary analysis, etc. \cite{zhu_15}.  

A signalling scheme with good JR/LPD/LPI capability has the following desirable attributes and functionality:
\begin{enumerate}
\item[A1.] \emph{Gaussianity:} This involves making the modulated signals featureless and indistinguishable from naturally occurring thermal white noise.  This requires the signal to be Gaussian distributed and have  a flat power spectral density \cite{marcio_may17, micha_nov10}.
\item[A2.] \emph{Ability to operate under receiver's noise floor:}  This is achieved using DSSS-based techniques to spread the narrowband message signal over a wide bandwidth, reducing required transmit power per channel symbol. These techniques are also widely used for anti-jamming wireless communication \cite{torr_18}.  
\item[A3.] \emph{Physical layer security:}  This involves complementing the overall security of the system through techniques adopted to secure transmission at the physical layer.  Traditionally, data security is achieved using encryption/decryption algorithms and is implemented at upper network layers 
(e.g. data link layer).  These algorithms normally encrypt only the information (message) component of the transmission.  This creates several vulnerabilities such as exposing details about the traffic flow and the source/destination addresses that could be used to launch denial of service attacks \cite{hoosh_dec17}. 
\item[A4.] \emph{Non-repetition:} This requires to use non-repetitive spreading sequences to reduce cyclic, temporal/spectral features \cite{kad_16}.  
\item[A5.] \emph{Uncoordinated synchronisation:} This requires a mechanism to acquire the signal and perform frame synchronisation without the use of a shared key or a fixed unique word \cite{popper_00}.
\end{enumerate}
Using the desired attributes listed above as a basis, this paper uses ML techniques to develop a synchronisable communication scheme that is shown to encapsulate these featureless.  

The rest of this paper is organised as follows.  Section \ref{section_dsss_system} briefly describes the standard DSSS system and features of its signals.  The proposed ML-based scheme is then presented and its performance is evaluated in Section \ref{section_MLSS}.  Finally, concluding remarks are presented in Section \ref{section_conclu}.

\hfill
\section{Standard DSSS System}
\label{section_dsss_system}
\begin{figure}[h!]
	\centering\small
	\includegraphics[width=.45\textwidth]{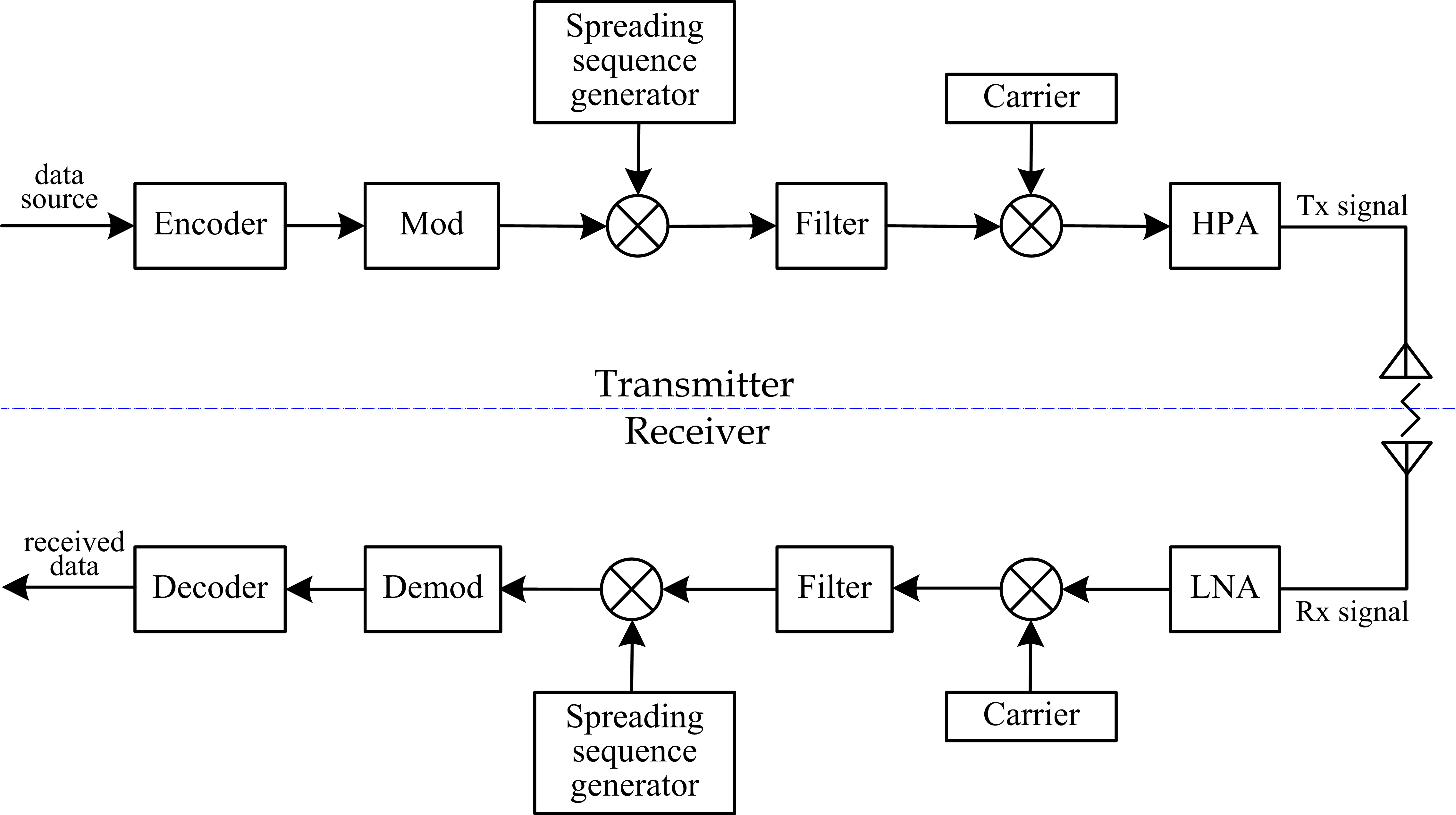}
	\caption{Block diagram of a standard DSSS system}
	\label{dsss}
\end{figure}

The block diagram of a standard DSSS system is depicted in Figure \ref{dsss}.  
In this system, the input source data is first encoded by an error correction code and then spread by a binary Pseudo-Noise (PN) sequence.  Spreading is performed by combining an $N$-length PN sequence with each encoded/modulated data bit.  Each symbol in the spread signal is referred to as a chip, and each bit can get a different sequence of chips.  The generated chips are then filtered, modulated (by a carrier) and then amplified before transmitting it over a channel.  Upon reception, the receiver despreads and decodes the wideband signal to retrieve the transmitted bits. Despreading is normally performed by correlating the received signal with the same PN sequence used by the transmitter in a synchronous manner. The output from the correlator is positive if the transmitted bit is one and negative if it is zero.  On retrieving the transmitted bit sequence, the receiver then searches for a pre-defined unique word in the bit stream to determine the start of each frame.      

Ideally, spreading sequences are required to have Autocorrelation Functions (ACF) with vanishing magnitudes, except at zero delay.  However, PN sequences tend to produce detectable spikes on its ACF due to its unique distribution.  M-sequences are one of the commonly used PN sequences \cite{torr_18}.  These sequences are typically generated using Linear Feedback Shift Registers (LFSR) in a deterministic manner and can have short periodic cycles if a large number of LFSRs are not used.  These sequences have been demonstrated to have relatively poor LPD properties due to its periodic nature and simple construction method \cite{burel_00}. Further, PN sequences have the potential to be easily predictable and hence can compromise security of the system. Some of the DSSS systems also rely on a shared secret key between the transmitter and the receiver to generate the random spreading sequence at the receiver.   However, from an implementation point of view,  establishing this key securely prior to every transmission is difficult and challenging \cite{popper_00}.

Spreading allows communication at low Signal-power to Noise-power Ratios (SNR), making it difficult for an unintended receiver to detect the signal.  The reduction in SNR is achieved by the amount of spreading used (processing gain). The total processing gain, $P_{G}$, achieved from a coded system with Binary Phase Shift Keying (BPSK)  is equal to the sum of waveform spreading gain, coding gain and gain due to code rate.

\begin{figure}[h!]
        \centering\small
        \includegraphics[width=.4\textwidth]{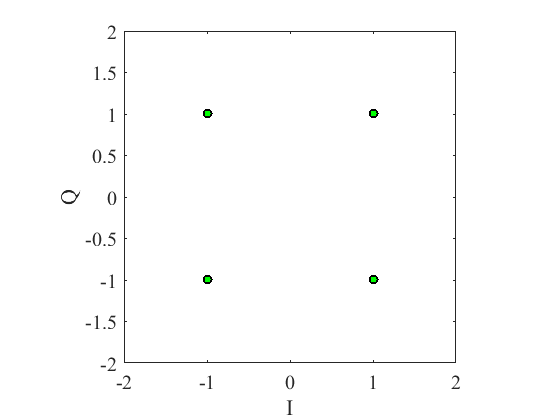}
		\caption{Signal constellation of the DSSS-PN (QPSK) transmit signal}
        \label{pn_sigset}
    \end{figure}

\begin{figure}[h!]
	\centering
	\subfigure[DSSS-PN signal]{%
		\label{pacf_DSSSPN}%
		\includegraphics[height=1.05in]{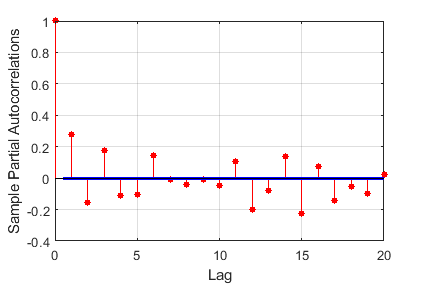}}%
	\qquad
	\subfigure[Gaussian noise]{%
		\label{pacf_noise}%
		\includegraphics[height=1.05in]{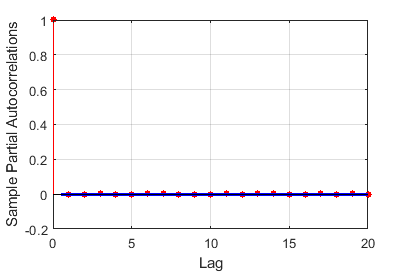}}%
	\caption{Partial autocorrelation functions}
	\label{pacf_PN}
\end{figure}

Figure \ref{pn_sigset} shows the signal constellation of a PN-based DSSS transmit signal that uses two spreading sequences to spread data bits 1 and 0.  This signal is modulated with Quadrature Phase Shift Keying (QPSK) and uses a spreading factor of 64.  
The signal constellation of the QPSK-modulated DSSS-PN system shows four distinct, equally-spaced points on its signal set. 
Figure \ref{pacf_PN} show Partial Autocorrelation Functions (PACF) \cite{box_94} of the DSSS-PN signal and Gaussian noise sequence respectively.  
Compared to standard ACF,   PACF provides a more clear picture of the correlations in the time signal.  
The PACF for the Gaussian noise case shows no correlation in time.  Its PACF function has a value of one at the zero-lag with all other values within the confidence band, indicating statistically insignificant correlation.  However, PACF for the DSSS-PN case shows strong correlated components demonstrating periodic use of the same spreading sequence.  It was also observed that the power spectrum of the Gaussian distributed noise is flatter than that of the DSSS-PN signal.

The following section describes a ML-based scheme for generating featureless spread signals.

\hfill
\section{ML-based Spread Spectrum (MLSS)}
\label{section_MLSS}

Machine learning (ML) is a branch of artificial intelligence that gives systems the ability to automatically learn and improve performance from experience without being explicitly programmed \cite{wang_nov17}.  During the last few years, application of ML techniques for solving problems in wireless communication has gained tremendous popularity in the research community \cite{jian_17}.  ML techniques have been applied to develop detection and equalisation schemes, channel estimation methods, channel decoding and demodulation schemes, etc. \cite{dorn_18}.  

\hfill
\subsection{The Autoencoder Concept}
An interesting ML-based concept for the physical layer has been proposed in \cite{tom_dec17}.  This concept, referred to as the \emph{Autoencoder}, views the complete communication system as an end-to-end optimisation problem and tries to reconstruct the transmitted message at the receiver output.  This is a deviation from the conventional design approach where signal processing modules (encoder, modulator, channel estimator, demodulator, etc.) are individually optimised using communication theory for a known channel model.  In contrast, the Autoencoder concept promises a method that can be used to develop waveforms for complex environments with unknown channel models.  Further, it also shows a way to develop advanced communication systems with increased ability to adapt and optimise to a dynamically changing environment using real-time learning. The Autoencoder architecture used in \cite{tom_dec17} is a fully-connected feed-forward neural network similar to the network shown in Figure \ref{net2}. This network uses multiple hidden layers for deep learning. The channel is modelled as an Additive White Gaussian Noise (AWGN) channel.  For a message block size of $k$, the network takes a vector of length $2^{k}$ as the input. 
The layer in the centre of the network is depicted as the \emph{channel} with $n$ neurons.  The $n$ analogue inputs that feed this layer denote the transmitted $n$ symbols.  Once the network is trained to minimise the error rate, the receiver (referred to as decoder) attempts to classify the corrupted transmitted message. 

With its many advantages, there are several challenges that have to be addressed before the Autoencoder can be used for more useful scenarios.  Some of these challenges are listed below.
\begin{enumerate}
\item[(i)] The implementation of Autoencoder is restricted by the number of inputs to the network.  For example, a message length of $k = 128$ bits would require an input vector of $2^{128}$ elements. Each element goes through a unique neuron in the input layer.  Implementing a network with such a large number of input neurons is practically impossible. 
\item[(ii)] The deep-learning architecture with multiple hidden layers also adds to the computational intensity of the encoding and decoding operations.  However, it should be noted that, depending on the message length and channel complexity, use of a deep-learning architecture may be required to achieve an acceptable level of performance.  
\end{enumerate}
Taking these issues and challenges into consideration, this section proposes two single hidden-layer feed-forward neural architectures for generating featureless Gaussian signals.

\hfill
\subsection{Proposed MLSS Architecture}

\begin{figure}[h!]
        \centering\small
        \includegraphics[width=.48\textwidth]{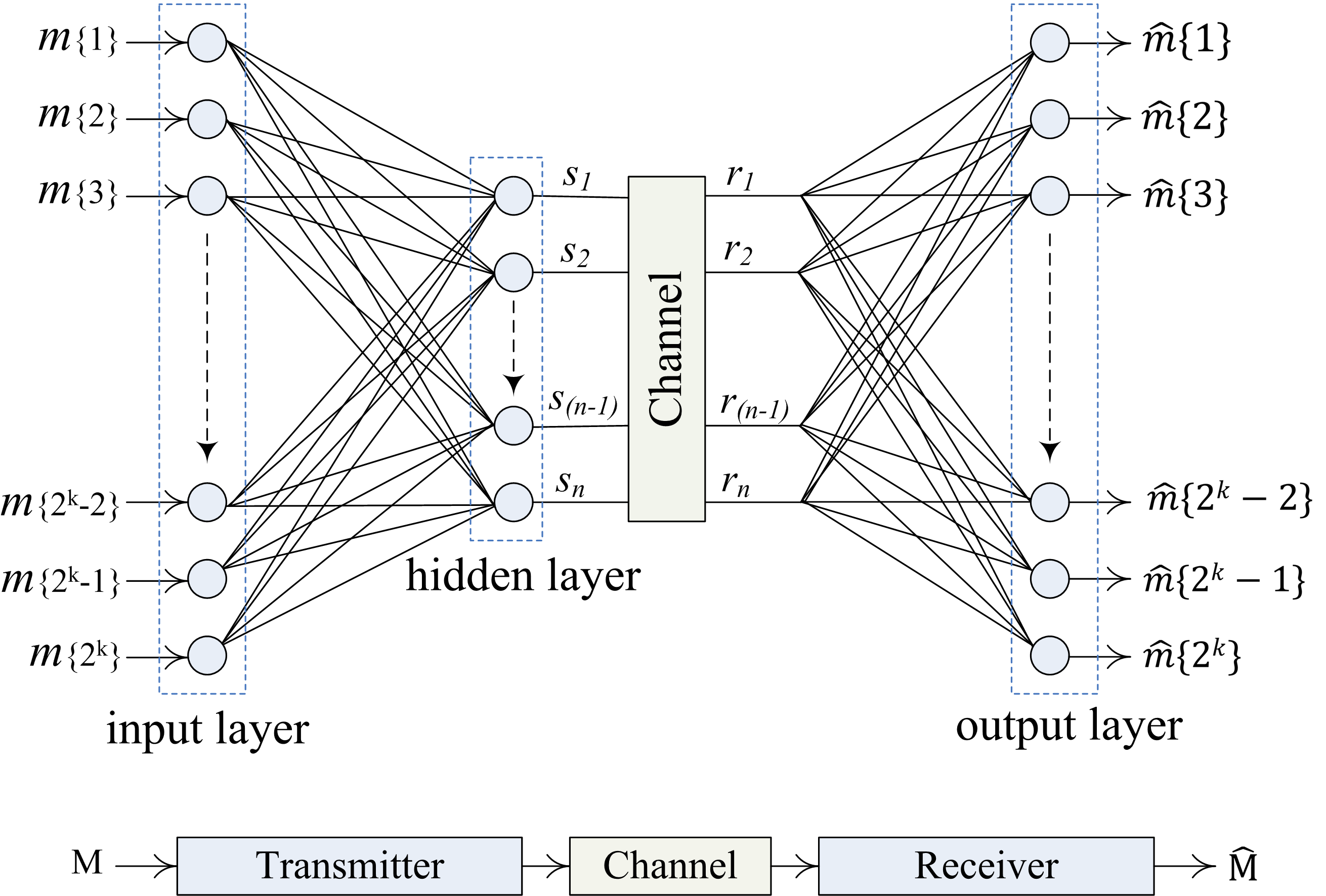}
        \caption{Single hidden layer neural network}
        \label{net2}
    \end{figure}

The proposed network is shown in Figure \ref{net2}.    
In the standard DSSS system, each data bit is spread using an $N$-length PN sequence.  In contrast, the proposed network can be viewed as an ML-based spreading network that spreads data, block by block.  Similar to the Autoencoder described earlier, this network takes an input vector of length $2^{k}$ elements with one element set to one and others set to negative one.  Once trained, this network spreads each $k$-bit block of data to $k$x$N$ analogue channel chips (symbols).  Assuming the length of each block is sufficiently large, it will be extremely difficult to despread and retrieve message blocks without knowing the network's trained weights \cite{dowlin_16}. This feature enhances transmission security of the system.

For the proposed system, the energy per chip to noise power spectral density ratio $E_{c}/N_{0}$, also referred to as SNR in this paper can be expressed as,
\begin{equation}\label{snrEq}
\text{SNR[dB]} = \frac{E_{b}}{N_{0}}\text{[dB]} - 10\log_{10}(\text{N}) 
\end{equation}
where $E_{b}/N_{0}$ is the energy per information bit to noise power spectral density ratio and N = $n/k$, where N is the spreading factor, $k$ is length of data block and $n$ is the number of channel symbols.   

The proposed network uses the \emph{one-step secant backpropagation algorithm} \cite{trainoss} to minimise the loss function and find the optimal weights of the network.  Softmax and cross-entropy functions are used for the output layer activation and the training loss functions respectively.  Softmax function calculates output ($\hat{m_{i}}$)  using the following expression,
\begin{equation}\label{softmax}
\hat{m_{i}} = \frac{e^{\hat{z}_{i}}}{\sum_{i=1}^{2^k}e^{\hat{z_{j}}}},  \text{  for } j = 1, ..., 2^{k}
\end{equation}  
where $\hat{z_{i}}$ are the inputs to the activation function.  The cross-entropy loss, $L$, is calculated using,
\begin{equation}\label{crossEntro}
L = \frac{1}{2^k}\sum_{i=1}^{2^k}l_{i}, 
\end{equation} 
where,
\begin{equation}\label{crossEntP}
l_{i} = -p_{i}\log(\hat{p_{i}}), \text{  for } i = 1, ..., 2^{k}
\end{equation} 
and
\begin{align}\label{crossEntMaxMin}
\hat{p_{i}}& = \max(\min(\hat{m_{i}},1-\epsilon),\epsilon), \\
p_{i} &= \max(\min(m_{i},1),0), 
\end{align} 
where $m_{i}$ and $\hat{m_{i}}$ represent the input and output values of the network and $\epsilon$ is the smallest number on the machine on which training is being performed. The network determines the most likely transmitted message by finding the index, $i$, of the highest probability value in the network output vector $\{\hat{m_{i}}\}$.  The performance of the proposed network is investigated in the following section.

\hfill
\subsubsection{Performance Investigation}
\begin{figure}[h!]
	\centering\small
	\includegraphics[width=.5\textwidth]{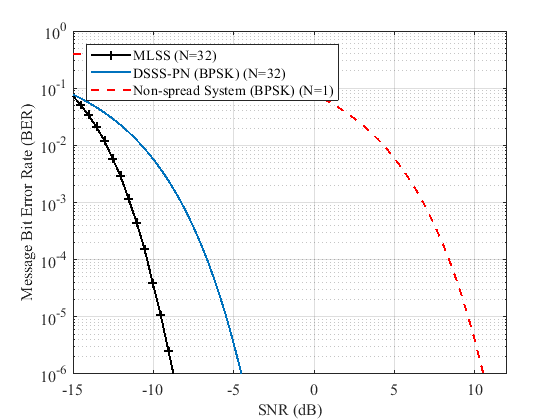}
	\caption{SNR performance of MLSS system, N=32}
	\label{mlss32_SNR}
\end{figure}

\begin{figure}[h!]
	\centering\small
	\includegraphics[width=.5\textwidth]{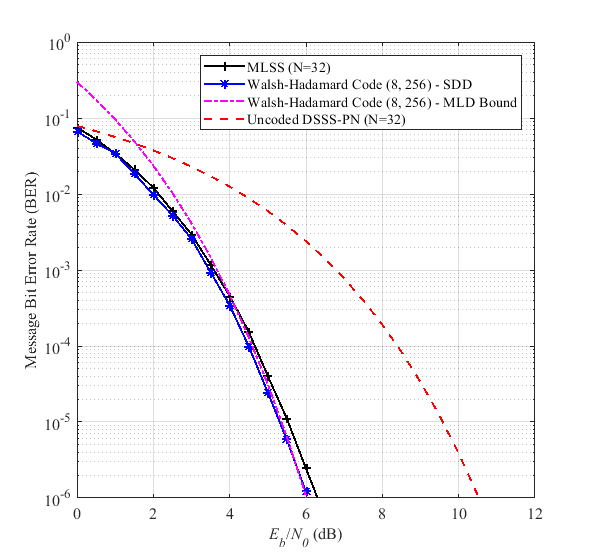}
	\caption{BER performance of MLSS system, N=32}
	\label{mlss32_BER}
\end{figure}

Figures \ref{mlss32_SNR} and \ref{mlss32_BER} show  Bit Error Rate (BER) performance against SNR and $E_{b}/N_{0}$ respectively for the system with an input message block size, $k$ = 8 bits and a spreading factor, N = 32. The network uses $2^{8}$ (i.e. 256) inputs/outputs to denote individual message blocks.  The number of outputs from the hidden layer (i.e. transmitter output) is 256 analogue values. These values are modulated and transmitted by mapping the generated values on to a one-dimensional signal constellation. The performance of the MLSS system is also compared with the BPSK-modulated DSSS-PN system. The network is trained on a noise-free channel.  The network settings used during training and performance evaluations are shown in Table \ref{mlss_sets}. 
\begin{table}[h!]
	\centering
	\caption{MLSS network settings}
	\label{mlss_sets}
	\begin{tabular}{lc}
		\hline
		Parameter                          & Setting                         \\
		\hline
		Training algorithm                 & One-step secant backpropagation \\
		Number of hidden layers            & 1                               \\
		Number of inputs - Input layer     & 256                             \\
		Number of neurons - Hidden layer   & 256                             \\
		Number of neurons - Output layer   & 256                             \\
		Activation function - Hidden layer & Linear                          \\
		Activation function - Output layer & Softmax       					 \\
		Loss function                      & Cross-entropy             		 \\
		Channel                            & AWGN                            \\
		Trained SNR                        & $\infty$ (Noise-free)           \\
		Weight/bias initialisation         & Uniformly distributed $\{-1,1\}$ 	\\
		Number of trained weights          & 131584 						\\
		\hline                   
	\end{tabular}
\end{table}

Figure \ref{mlss32_SNR} shows that, to achieve a target BER of $10^{-6}$, the MLSS system can operate 4.3dB below the corresponding operating SNR required for the DSSS-PN system. It should be noted that this gain in SNR is achieved by keeping the spreading factor (and bandwidth expansion) the same for both systems.  The BER performance shown in Figure \ref{mlss32_BER} indicates that the trained system (i.e. trained on a noise-free channel) performs close to the theoretical Maximum-Likelihood Decoding (MLD) limit and optimum Soft-Decision Decoding (SDD) performance of the Walsh-Hadamard code ($n$=256, $k$=8, $d_{min}$=128, $T$=63). Here, the parameters $d_{min}$ and $T$ are the minimum Hamming distance and error correcting capability of the code.

\begin{figure}[h!]
	\centering
	\subfigure[Signal constellation]{%
		\label{sigset_mlss}%
		\includegraphics[height=2.5in]{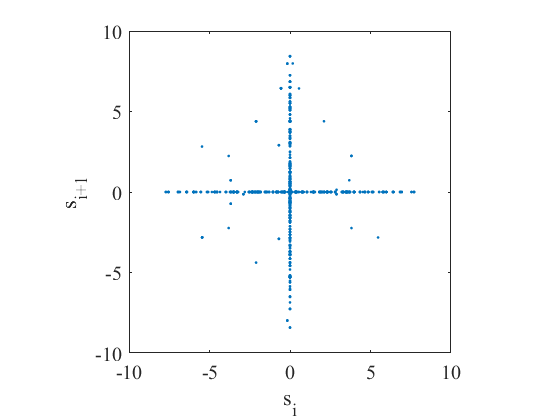}}%
	\qquad
	\subfigure[Autocorrelation]{%
		\label{auto_mlss}%
		\includegraphics[height=2.1in]{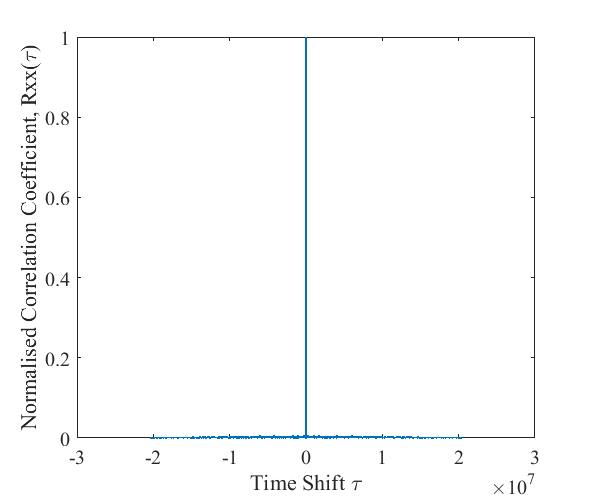}}%
	\caption{Features of the MLSS signal trained in noise-free environment}
	\label{mlss_noNoise_features}
\end{figure}

The signal constellation and ACF of the MLSS signal are shown in Figure \ref{mlss_noNoise_features}.  
The ACF plot indicates no repetitive patterns in the signal sequence.  Each point in the signal constellation is represented by a non-overlapping signal pair ($s_{i}$, $s_{(i+1)}$).  The plotted constellation shows a non-Gaussian distribution as a large number of individual signal values converge on zero.  Signals with more noise-like features with Gaussian distribution can be generated by training the network on an AWGN channel, as demonstrated in the following section.  The training process finds an optimal set of weights by adjusting its signal distribution to that of the channel.  

\hfill
\subsection{Computationally Efficient Architecture}
\begin{figure}[h!]
	\centering\small
	\includegraphics[width=.5\textwidth]{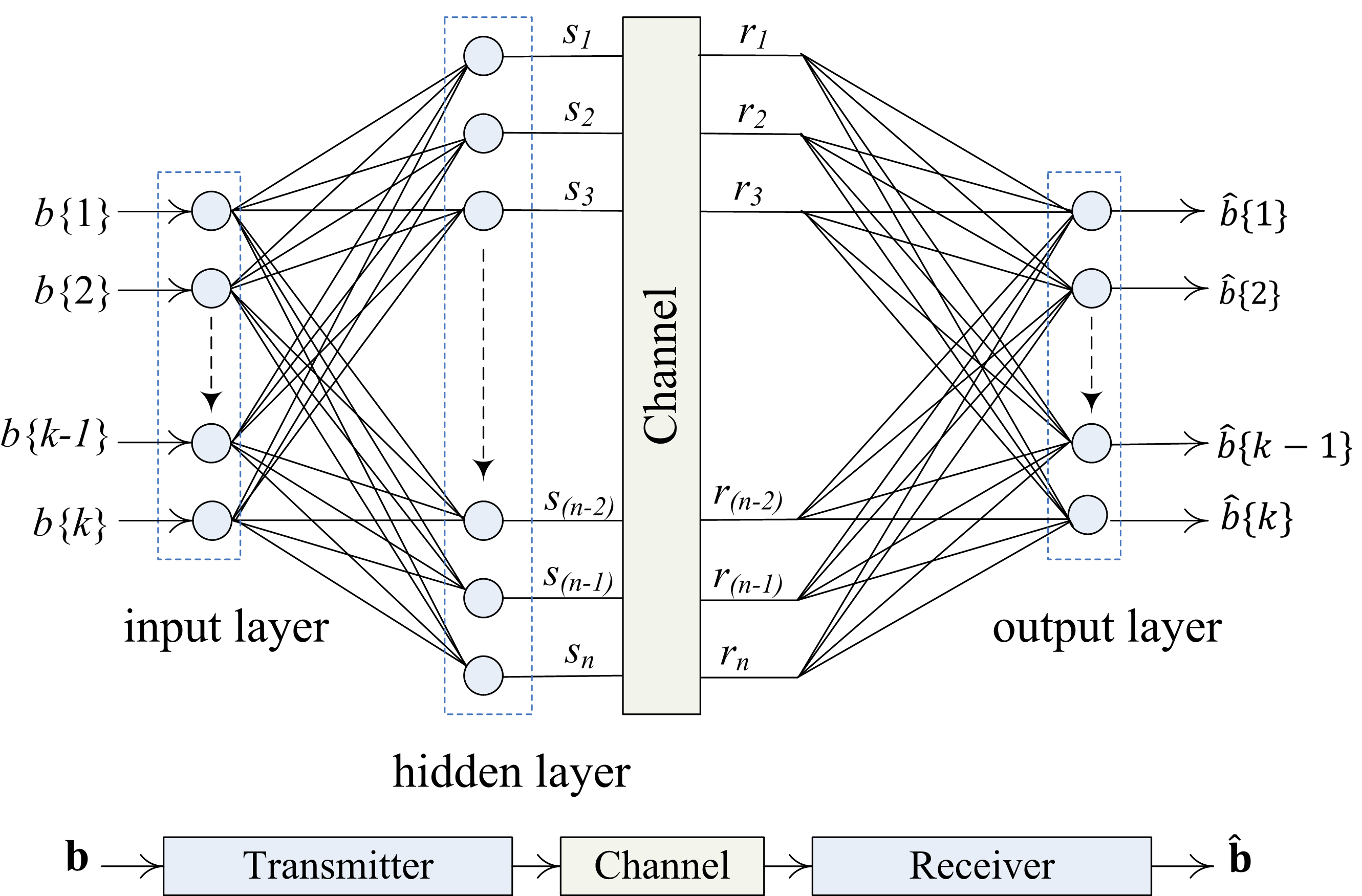}
	\caption{Single hidden layer neural network with direct inputs}
	\label{net1}
\end{figure}

\
The architecture proposed in the previous section require $2^{k}$ neurons on its input layer for a message block size of $k$ bits.  As a result, the complexity of the network increases exponentially with the size of the message block, hindering its implementation for more useful cases.  A computationally more efficient network architecture is proposed in Figure \ref{net1}.  
This network uses $k$ neurons on its input and output layers.  The input vector is represented by $\{b_{i}\}$, $i$ = 1, 2, ..., $k$, where $b_{i}$ denotes $i^{th}$ bit in the message block. Before feeding this vector to the network, the elements zero and one in the vector are first mapped to negative one and one respectively.  The number of weights in the network is 133152.  The network is trained on an AWGN channel at sufficiently low SNR to generate Gaussian distributed analogue symbols from its hidden layer. It was observed that the proposed network architecture with reduced input dimension performs close to the uncoded system ($E_{b}/N_{0}$ vs BER) and does not give any coding gain. To compensate for this, the fully trained network is concatenated with a suitable error correction coding scheme to achieve an acceptable level of performance.  This network setup provides a good tradeoff between complexity and performance.  

The proposed network is trained using the same backpropagation algorithm used for the previous architecture.  The Elliot Symmetric Sigmoid (ESS) function \cite{elliot_93} is used as the activation function at the output layer.  Unlike many common activation functions, ESS does not use exponentials and hence is computationally very efficient. However, compared to other Sigmoid functions, ESS may require more training iterations to achieve the same accuracy.  ESS function is defined by,
\begin{equation}\label{ess}
\hat{m_{j}} = \frac{1}{1+|z_{j}|},  \text{  for } j = 1, ..., 2^{k}
\end{equation}  
Mean Absolute Error (MAE) function is used as the loss function.  MAE for the proposed network architecture is defined by,
\begin{equation}\label{mae}
\text{MAE} = \frac{1}{2^k}\sum_{i=1}^{2^k} |m_{i}-\hat{m_{i}}| 
\end{equation}
The performance of the described network is investigated in the following section.

\hfill
\subsubsection{Performance Investigation}
\begin{figure}[h!]
	\centering
	\subfigure[Signal constellation]{%
		\label{sigset_mlss6dB}%
		\includegraphics[height=2.7in]{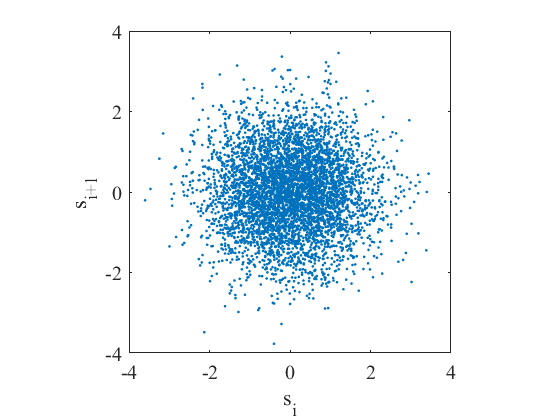}}%
		\qquad
	\subfigure[Autocorrelation]{%
		\label{auto_mlss6dB}%
		\includegraphics[height=1.2in]{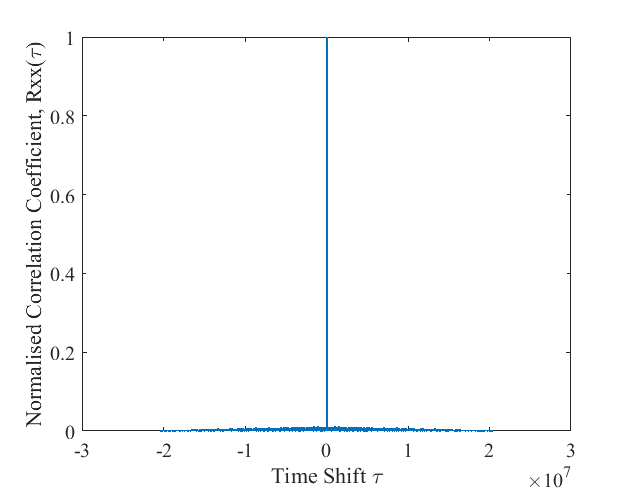}}%
			\qquad
	\subfigure[Partial autocorrelation]{%
		\label{pacf_mlss6dB}%
		\includegraphics[height=1.2in]{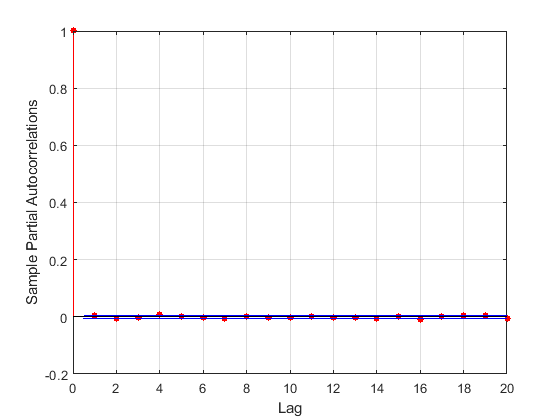}}%
	\caption{Features of the noise-free MLSS signal (trained at $E_{b}/N_{0}$ = 6dB,  SNR = -12dB)}
	\label{mlss_features}
\end{figure}

\begin{figure}[h!]
	\centering\small
	\includegraphics[width=.54\textwidth]{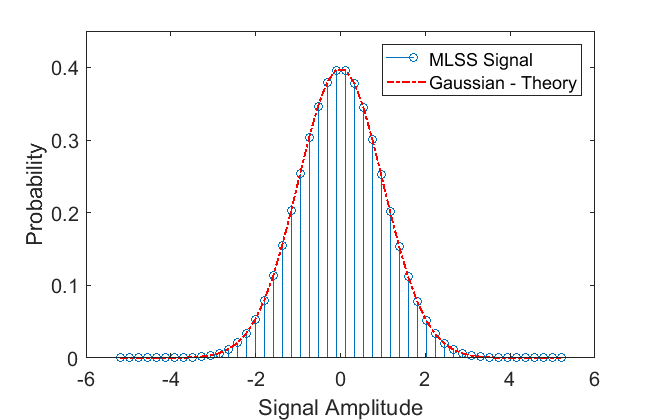}
	\caption{Distribution of the noise-free MLSS signal}
	\label{pdf_mlss6dB}%
\end{figure}

Performance of the network is investigated using a message block size of 32 bits with a 2048-neuron hidden layer. This provides a processing gain of 64.  The network is trained at an SNR of -12dB (corresponding to an $E_{b}/N_{0}$ of 6dB).  Figure \ref{mlss_features} shows the signal constellation and autocorrelation functions of the noise-free MLSS signal generated from the trained network.  The signal constellation shows a noise-like signal.  The autocorrelation functions obtained indicate no correlation in the signal.  
The distribution of the noise-free chips generated from the trained network is compared with the Standard Gaussian Distribution (SGD) in Figure \ref{pdf_mlss6dB} and Table \ref{moments}.  These results are obtained using $2048$x$10^{5}$ analogue values.   The probability distribution of the MLSS signal (Figure \ref{pdf_mlss6dB}) closely matches the theoretical Gaussian distribution.  
\begin{table}[]
	\centering
	\caption{Moment Comparison of MLSS with Theory}
	\label{moments}
	\begin{tabular}{lll}
		\hline
		\textbf{Moment}    & \textbf{SGD} & \textbf{MLSS} \\
		\hline
		$1^{st}$          & 0                              & 0.00        \\
		$2^{nd}$          & 1                              & 1.00        \\
		$3^{rd}$          & 0                              & 0.00       \\
		$4^{th}$          & 3                              & 2.98        \\
		$5^{th}$          & 0                              & 0.00       \\
		$6^{th}$          & 15                             & 14.73       \\
		$7^{th}$          & 0                              & -0.05       \\
		\hline 
	\end{tabular}
\end{table}
The Gaussianity of the noise-free MLSS signal is also validated by computing higher order moments.  
The estimated moments (Table \ref{moments}) indicate an acceptable level of similarity with the theoretical Gaussian distribution.

\begin{figure}[h!]
	\centering\small
	\includegraphics[width=.53\textwidth]{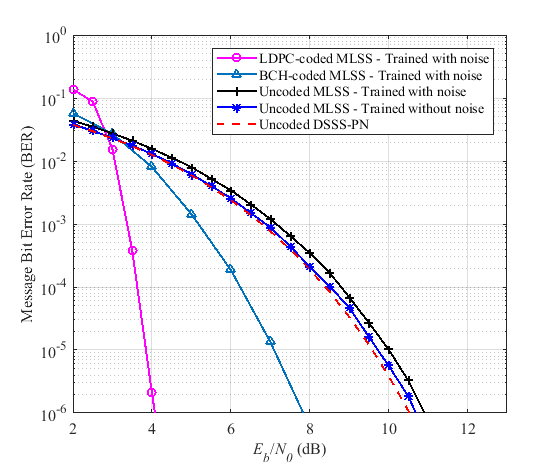}
	\caption{BER performance of direct input MLSS system (Trained at SNR = -12dB), N=64 [$k$=32, $n$=2048] }
	\label{mlss64_BER}
\end{figure}

The BER performance of the network is shown in Figure \ref{mlss64_BER}.  This figure presents results for the uncoded and coded MLSS systems.  Two coded systems are developed by concatenating error correction codes with the network trained with noise.  The first coded system uses an extended BCH (31,11) code, while the second uses a 1/2-rate LDPC (1944, 972) code.
Performance results obtained show that the LDPC-coded MLSS system gives a total processing gain of approximately 27.5dB.  This system achieves a BER of $10^{-6}$ at an SNR of -17dB ($E_{b}/N_{0}$ = 4dB). Soft-decision decoding was used for both BCH and LDPC decoding. Noise variance required for computing Log-Likelihood Ratios (LLR) for the LDPC decoder were estimated using the noisy chips acquired at the receiver. 

\hfill
\subsection{Implementation of the MLSS Signal}
\label{section_implement}

\begin{figure}[h!]
	\centering\small
	\includegraphics[width=.48\textwidth]{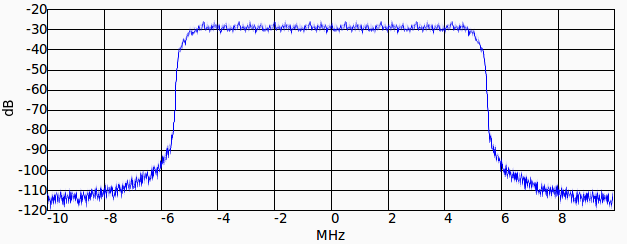}
	\caption{Spectrum of the implemented LDPC-coded MLSS signal}
	\label{psd_gnu}
\end{figure}

A Software-Defined Radio (SDR) transmitter for the proposed scheme is implemented on MATLAB\textsuperscript{TM}/GNU-Radio/Ettus\textsuperscript{TM}-USRP. The  LDPC-coded MLSS signal generated from the trained network is interpolated, and filtered using a Root-Raised-Cosine filter.  The bandlimited signal (confined to a bandwidth of 10 MHz) is then transferred to an Ettus Universal Software Radio Peripheral (USRP).  At the USRP, the MLSS signal is upconverted to a carrier Radio Frequency (RF) of 1 GHz and then transmitted from its transmit port and received back to the SDR through its receive port.  The power spectrum of the bandlimitted received signal is shown in Figure \ref{psd_gnu}.
The captured signals show a flat power spectrum, indicating the Gaussian noise-like feature of the MLSS signal.  With the estimated processing gain of 27.5dB and Gaussian distributed nature of the signal, the received signals can easily blend into the receiver's background noise, making it indistinguishable.

\subsection{Synchronisation of the MLSS System}
\label{section_sync}
In the DSSS-based systems, the spreading sequences are known to both transmitter and receiver.  The key objective of synchronisation is to align the spreading sequences used by the transmitter with that used at the receiver. The receiver normally initiates generating sequences using a secret key known to the transmitter. Synchronisation is performed by correlating the received signal with the receiver-generated sequence.  Synchronisation between the spreading and despreading sequence is established when the correlation magnitude exceeds a threshold. This process is referred to as sequence acquisition.  Once the two sequences are aligned, the receiver then locates the start of a pre-defined unique word (or preamble) in the despread bit stream to determine the start of the frame.  This process is referred to as frame synchronisation. 

\begin{figure}[h!]
        \centering\small
        \includegraphics[width=.48\textwidth]{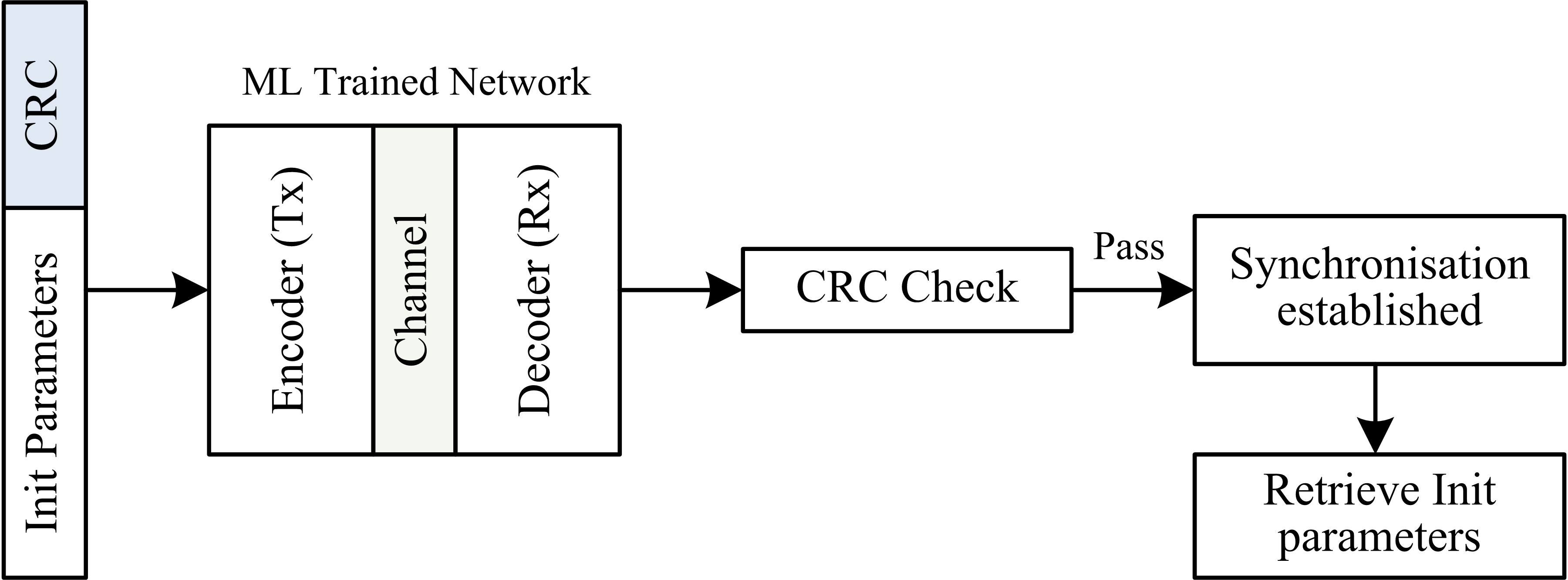}
        \caption{MLSS synchronisation}
        \label{sync}
    \end{figure}

The synchronisation technique proposed for the MLSS system is depicted in Figure \ref{sync}.    
The proposed technique carries out sequence acquisition and frame synchronisation jointly by encoding the network message block with a Cyclic Redundancy Check (CRC) code.  At the transmitter, the trained network encoder maps the CRC-coded block to a set of channel symbols for transmission.  At the receiver, the trained network decoder acquires a stream of channel symbols, decodes and then performs a CRC check on it. Synchronisation is established when the CRC check passes and the corresponding time offset is used to determine start of future frames. CRC-coded message blocks can also be transmitted at regular intervals to maintain synchronisation. It is assumed that the message blocks are encrypted and preferably interleaved across multiple blocks to randomise and minimise data repetition at the input of network encoder. The \emph{Init Parameters} field in the message block can be filled with dummy random values or if required, used to deliver vital system parameters to the receiver.  MLSS can also be used to convey initialisation parameters for other noise signalling schemes, such as Chaos-based signalling techniques \cite{kad_16} that are difficult to synchronise.



\hfill
\section{Conclusion}
\label{section_conclu}
Motivated by the poor LPD/LPI capability of the standard DSSS-PN systems, this paper proposes a scheme that is shown to generate synchronisable featureless signals using machine learning. Error performance of this scheme and the characteristics of the generated signals are investigated and compared with the standard DSSS signals.  The results obtained show that the proposed scheme generates uncorrelated spread signals with good JR/LPD/LPI features. Gaussianity of the signals are analysed using its distribution and higher-order moments. Autocorrelation functions are also plotted to identify any repetitive patterns in the signal.  These tests reveal no identifiable features.  However, it should be noted that the transmit filter, HPA and RF modulation can add some detectable features to the signal and more advanced methods such as  cyclostationary-based techniques \cite{gard_92, vlok_14} would be required to identify any possible existence of such features in the signal.

\hfill
\section*{Acknowledgment}
The author would like to thank Daniel Salmond, Perry Blackmore, Weimin Zhang, Jeewani Kodithuwakkuge and Raymee Chau for many helpful discussions and suggestions.         

\hfill


\begin{thebibliography}{1}
\bibliographystyle{IEEEtran}

\bibitem{tom_dec17}
T. O'Shea, J. Hoydis, An Introduction to Deep Learning for the Physical Layer,  IEEE Transactions on Cognitive Communications and Networking, pp. 563-575, Vol. 3, No. 4, Dec. 2017. 

\bibitem{wang_nov17}
T. Wang, et. al. Deep Learning for Wireless Physical Layer: Opportunities and Challenges, China Communication, pp. 92-111, Nov. 2017. 

\bibitem{zhu_15}
Z. Zhu and A. K, Nandi, Automatic Modulation Classification: Principles, Algorithms and Applications, Wiley, UK, 2015


\bibitem{marcio_may17}
M. Eisencraft, et. al., White Gaussian Chaos, IEEE Communication Letters, pp. 1719-1722, Vol. 8, Issue 8,  Aug. 2017. 

\bibitem{micha_nov10}
A. J. Michaels, D. B. Chester, Efficient and Flexible Chaotic Communication Waveform Family,  IEEE Military Communications Conference, California, USA, Nov. 2010.


\bibitem{torr_18}
D. Torrieri, Principles of Spread-Spectrum Communication Systems, 4th Ed., Springer, USA, 2018.

\bibitem{hoosh_dec17}
R. Hooshmand and M. R. Aref,  Efficient Polar Code-Based Physical Layer Encryption Scheme, IEEE Wireless Communications Letters, pp. 710-713, Vol. 6, Issue 6, Dec. 2017. 

\bibitem{kad_16}
G. Kaddoum, Wireless Chaos-Based Communication Systems: A Comprehensive Survey,  IEEE Access, pp. 2621-2648, Vol. 4, May 2016.

\bibitem{popper_00}
C. Popper, et. al., Anti-jamming broadcast communication using uncoordinated spread spectrum techniques, IEEE Journal on Selected Areas in Communications, pp. 703-715,  Vol. 28, Issue 5, June 2010. 

\bibitem{burel_00}
G. Burel, et. al., Blind Estimation of the Pseudo-random of a Direct Sequence Spread Spectrum Signal, IEEE Military Communication Conference, USA, Oct. 2000.

\bibitem{box_94}
G. E. Box, et. al., Time Series Analysis: Forecasting and Control, 3rd ed., Englewood Cliffs, NJ: Prentice Hall, 1994.





\bibitem{jian_17}
C. Jiang,  Machine Learning Paradigms for Next-Generation Wireless Networks, IEEE Wireless Communications, pp. 98-105, Vol. 24, Issue 2, Apr. 2017.


\bibitem{dorn_18}
S. Dorner, et. al., Deep Learning Based Communication Over the Air,   IEEE Journal of Selected Topics in Signal Processing, Vol. 12, Issue 1, Feb. 2018.   

\bibitem{dowlin_16}
N. Dowlin, et. al.,  CryptoNets : Applying Neural Networks to Encrypted Data with High Throughput and Accuracy, International Conference on Machine Learning, New York, USA, June 2016. 

\bibitem{trainoss}
R. Battiti, First and Second Order Methods for Learning: Between Steepest Descent and Newton’s Method, Neural Computation, pp. 141–166, Vol. 4, No. 2, 1992. 

\bibitem{elliot_93}
D. L. Elliott, A Better Activation Function for Artificial Neural Networks, University of Maryland. Systems Research Center, 1993.

\bibitem{gard_92}
W. A. Gardner and C. M. Spooner, Signal Interception Performance Advantages of Cyclic-feature Detectors, IEEE Transactions on Communications, Vol. 40, Issue 1, pp. 149-159, Jan. 1992.

\bibitem{vlok_14}
J. D. Vlok and J. C. Olivier, Blind Sequence-length Estimation of Low-SNR Cyclostationary Sequences, IET Communications, Vol. 8, No. 9, pp. 1578-1588, Jun. 2014.






\end{thebibliography}
\end{document}